\documentclass[%
 reprint,
 amsmath,amssymb,
 aps,
 pra,
 superscriptaddress
]{revtex4-2}
\usepackage{graphicx}
\usepackage[T1]{fontenc}
\usepackage{dcolumn}
\usepackage{bm}
\usepackage{lipsum}
\usepackage{braket}
\usepackage{siunitx}
\usepackage{epigraph}
\usepackage{todonotes}
\setcitestyle{super} 
\usepackage{hyperref}

\newcommand{\sissis}{%
  \tikz[baseline=-0.6ex, x=1em, y=1em, yshift=-0.7pt]{%
    \def\s{0.18}   
    \def\sep{0.6} 
    \def\lead{0.35}

    \draw (-\lead,0) -- (\sep+\lead,0);
    
    \draw (-\s,-\s) -- (\s,\s);
    \draw (-\s,\s) -- (\s,-\s);

    \draw (\s,0) -- (\sep-\s,0);

    \begin{scope}[xshift=\sep em]
      \draw (-\s,-\s) -- (\s,\s);
      \draw (-\s,\s) -- (\s,-\s);
    \end{scope}

  }%
}

\newcommand{\sns}{%
  \tikz[
    baseline=-0.6ex,
    x=1em, y=1em,
    yshift=-0.7pt,
    line cap=round,
    line join=round
  ]{%
    \def\s{0.18}       
    \def\leadL{0.35}   
    \def\leadR{0.2}   
    \def\eps{0.01}     

    \draw (-\leadL,0) -- (-\s,0);

    \draw (-\s,-\s) rectangle (\s,\s);

    \draw[dash pattern=on 0.6pt off 0.6pt]
      (-\s,-\s) -- (\s-\eps,\s-\eps);
    \draw[dash pattern=on 0.6pt off 0.6pt]
      (-\s,\s) -- (\s-\eps,-\s+\eps);

    \draw (\s,0) -- (\s+\leadR,0);
  }%
}

\newcommand{\qubitacr}{HPQ}

\begin{document}

\title{Controlled Parity of Cooper Pair Tunneling in a Hybrid Superconducting Qubit}

\author{David Feldstein-Bofill}
\email{david.bofill@nbi.ku.dk}
\affiliation{Center for Quantum Devices, Niels Bohr Institute, University of Copenhagen, Denmark}
\affiliation{NNF Quantum Computing Programme, Niels Bohr Institute, University of Copenhagen, Denmark}

\author{Leo Uhre Jakobsen}
\affiliation{Center for Quantum Devices, Niels Bohr Institute, University of Copenhagen, Denmark}
\affiliation{NNF Quantum Computing Programme, Niels Bohr Institute, University of Copenhagen, Denmark}

\author{Ksenia Shagalov}
\affiliation{Center for Quantum Devices, Niels Bohr Institute, University of Copenhagen, Denmark}
\affiliation{NNF Quantum Computing Programme, Niels Bohr Institute, University of Copenhagen, Denmark}

\author{Zhenhai Sun}
\affiliation{Center for Quantum Devices, Niels Bohr Institute, University of Copenhagen, Denmark}
\affiliation{NNF Quantum Computing Programme, Niels Bohr Institute, University of Copenhagen, Denmark}

\author{Casper Wied}
\affiliation{Center for Quantum Devices, Niels Bohr Institute, University of Copenhagen, Denmark}
\affiliation{NNF Quantum Computing Programme, Niels Bohr Institute, University of Copenhagen, Denmark}

\author{Shikhar Singh}
\affiliation{Center for Quantum Devices, Niels Bohr Institute, University of Copenhagen, Denmark}
\affiliation{NNF Quantum Computing Programme, Niels Bohr Institute, University of Copenhagen, Denmark}

\author{Anders Kringhøj}
\affiliation{Center for Quantum Devices, Niels Bohr Institute, University of Copenhagen, Denmark}
\affiliation{NNF Quantum Computing Programme, Niels Bohr Institute, University of Copenhagen, Denmark}

\author{Jacob Hastrup}
\affiliation{Center for Quantum Devices, Niels Bohr Institute, University of Copenhagen, Denmark}
\affiliation{NNF Quantum Computing Programme, Niels Bohr Institute, University of Copenhagen, Denmark}

\author{Andr\'as Gyenis}
\affiliation{Department of Electrical, Computer \& Energy Engineering, University of Colorado Boulder, Boulder, CO 80309, USA}
\affiliation{Department of Physics, University of Colorado Boulder, Boulder, CO 80309, USA}

\author{Karsten Flensberg}
\affiliation{Center for Quantum Devices, Niels Bohr Institute, University of Copenhagen, Denmark}

\author{Svend Krøjer}
\affiliation{Center for Quantum Devices, Niels Bohr Institute, University of Copenhagen, Denmark}
\affiliation{NNF Quantum Computing Programme, Niels Bohr Institute, University of Copenhagen, Denmark}

\author{Morten Kjaergaard}
\email{mkjaergaard@nbi.ku.dk}
\affiliation{Center for Quantum Devices, Niels Bohr Institute, University of Copenhagen, Denmark}
\affiliation{NNF Quantum Computing Programme, Niels Bohr Institute, University of Copenhagen, Denmark}

\date{\today}

\maketitle

\section*{Abstract}
Superconducting quantum circuits derive their nonlinearity from the Josephson energy-phase relation. Besides the fundamental $\cos\phi$ term, this relation can also contain higher Fourier harmonics $\cos(k\phi)$ corresponding to correlated tunneling of $k$ Cooper pairs.
The parity of the dominant tunneling process, i.e.~whether an odd or even number of Cooper pairs tunnel, results in qualitatively different properties, and controlling this opens up a wide range of applications in superconducting technology. However, access to even-dominated regimes has remained challenging and has so far relied on complex multi-junction or all-hybrid architectures.
Here, we demonstrate a simple "harmonic parity qubit"; an element that combines two aluminum-oxide tunnel junctions in parallel to a gate-tunable InAs/Al nanowire junction forming a SQUID, and use spectroscopy versus flux to reconstruct its energy-phase relation at 85 gate voltage points. 
At half flux quantum, the odd harmonics of the Josephson potential can be suppressed by up to two orders of magnitude relative to the even harmonics, producing a double-well potential dominated by even harmonics with minima near $\pm\pi/2$. 
The ability to control harmonic parity enables supercurrent carried by pairs of Cooper pairs and provides a new building block for Fourier engineering in superconducting circuits.

\section*{Introduction}
\begin{figure}[t!]
\includegraphics{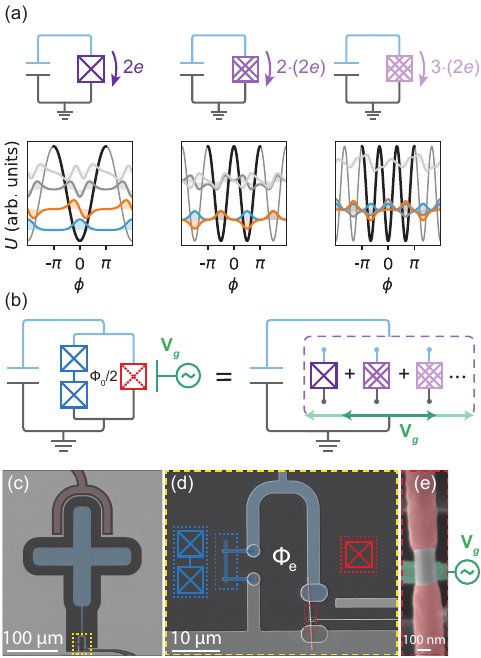}
\caption{Harmonic parity control in a hybrid superconducting qubit. (a) Superconducting qubit circuits based on Josephson elements that transfer one, two, or three Cooper pairs (left to right). The panels below show the corresponding Josephson potentials (black) and the wavefunctions of the two lowest states (blue and orange), and higher states (gray). Higher-order Cooper-pair tunneling gives rise to higher Josephson harmonics $\cos(k\phi)$, leading to different qubit properties. (b) Circuit realized in this work. The Josephson element is a SQUID with two SIS junctions in series on the left branch and a gate-tunable S-Sm-S nanowire junction on the right. The harmonic contents of the two SQUID branches interfere as a function of the magnetic flux $\Phi_\mathrm{e}$ threading the loop. At $\Phi_\mathrm{e}=\Phi_0/2$, the gate voltage controls whether an even or odd number of Cooper pairs dominates the tunneling process.
(c) False-colored scanning electron microscope image of the Harmonic Parity Qubit (HPQ) device. A $\lambda/4$ resonator (brown) is capacitively coupled to the qubit island (blue) for readout.
(d) Zoom of the SQUID loop showing the two SIS junctions (left) and the S-Sm-S junction (right).
(e) SEM of the InAs/Al nanowire junction. The junction is defined by removing $\sim\SI{200}{nm}$ of the epitaxial Al shell (red) covering the nanowire. The electrostatic gate (green) is patterned beneath the junction and counter-etched to leave a vacuum gap between the gate and the nanowire. \label{Fig1}}
\end{figure}

Josephson junctions are among the most fundamental building blocks in modern superconducting technology~\cite{blais2021circuit, kim2025josephson}. 
By coherently coupling macroscopic phase variables through the tunneling of Cooper pairs, they provide the nonlinearity underlying various classical and quantum devices from quantum-limited amplifiers to the leading superconducting qubit platforms~\cite{blais2021circuit,kjaergaard2020superconducting}.
In the most widely used superconductor-insulator-superconductor (SIS) tunnel junctions, the energy-phase relation is nearly sinusoidal, $U(\phi)\propto\cos\phi$, where $\phi$ is the phase drop across the junction, reflecting coherent tunneling of single Cooper pairs~\cite{josephson1962possible,golubov2004current}, illustrated in Fig.~\ref{Fig1}a (left panel).
This “$\cos(\phi)$-element” has been the workhorse of circuit quantum electrodynamics, enabling a large family of circuits including the weakly anharmonic transmon qubit \cite{koch2007charge}, the strongly anharmonic fluxonium qubit~\cite{manucharyan2009fluxonium}, complex multi-mode qubits\cite{bell2014protected, gladchenko2009superconducting,smith2020superconducting,smith2022magnifying,dodge2023hardware,gyenis2021experimental, brooks2013protected,hays2025non,gyenis2021moving} and a growing family of large-scale quantum processors and quantum-limited microwave devices~\cite{google2025quantum, kim2023evidence, krinner2022realizing, macklin2015near}. 

The Josephson effect, however, is not confined to single-pair tunneling. 
Whenever higher-order coherent transport processes become appreciable, the junction energy-phase relation acquires harmonics of the form of $c_k\cos(k\phi)$ with $k>1$, corresponding to the correlated tunneling of $k$ Cooper pairs with relative weight $c_k$, illustrated in Fig.~\ref{Fig1}a. 
Such multi-pair terms dramatically reshape the potential landscape and can qualitatively alter selection rules, wavefunction parity, and qubit noise sensitivity~\cite{pita2025novel, gyenis2021moving, hays2025non, larsen2020parity}. 
Tuning the weights of individual Fourier components ($c_k$) enables complete freedom in engineering the potential landscape~\cite{bozkurt2023double}, but at the cost of prohibaitively large experimental overhead. 
Instead, controlling the overall parity of the harmonic content, i.e.~the relative weight of odd and even harmonics, provides a manageable control knob that can tune to a regime dominated by even harmonics and thereby protecting Cooper pair parity by symmetry.

In particular, a tunable, $\pi$-periodic Josephson element dominated by $\cos(2\phi)$ is especially compelling.
The suppression of the fundamental Josephson harmonic results in tunneling dominated by pairs of Cooper pairs and enforces the conservation of Cooper-pair parity on the superconducting island, as illustrated in Fig.~\ref{Fig1}a (middle panel).
This symmetry is the central resource for intrinsically protected qubits, including the $\cos(2\phi)$ qubit based on the rhombus~\cite{bell2014protected, gladchenko2009superconducting, banszerus2025hybrid, sanchez2026revisiting,zhurbina2026coherence}, protected qubits based on the kinetic interference cotunneling element~\cite{smith2020superconducting,smith2022magnifying,dodge2023hardware,hays2025non,nguyen2025superconducting,roverc2026experimental}, parity-protected qubits based on S-Sm-S or twisted van-der-Waals junctions~\cite{larsen2020parity, liu2025strongly, brosco2024superconducting}, and related $0$-$\pi$~\cite{gyenis2021experimental, brooks2013protected} architectures, where a near-degenerate ground-state manifold is expected to be exponentially insensitive to key local noise channels~\cite{gyenis2021moving, messelot2026coherence}. 
In these schemes, the ability to design and control the balance among $\cos(\phi)$, $\cos(2\phi)$, and higher-order terms will ultimately set the level of protection and the viable operating space.

Recently, measurable higher harmonics have been detected even in conventional Al/AlOx/Al tunnel junctions, reviving interest in Josephson elements with higher Josephson harmonics. 
These studies report $\cos(2\phi)$ contributions at the single percent level; however, the harmonic content is still largely set by fabrication and circuit layout, rather than serving as a tunable circuit parameter~\cite{willsch2024observation, kim2025emergent}.
Alternatively, in hybrid superconductor-semiconductor-superconductor (S-Sm-S) junctions, Andreev bound states of high transparency generate an intrinsically nonsinusoidal potential with a sizeable harmonic content that can be tuned via an electrostatic gate~\cite{larsen2020parity,kringhoj2018anharmonicity,kringhoj2020suppressed,bargerbos2020observation,fatemi2025nonlinearity}.
This platform has been used to realize all-hybrid S-Sm-S junction interferometers, demonstrating transport and qubit characteristics consistent with tunneling dominated by pairs of Cooper pairs~\cite{larsen2020parity,ciaccia2024charge}. 
Despite this progress, the field still lacks a compact element in which the Josephson harmonic parity can be reliably tuned by a single external control knob, allowing deliberate control of the energy-phase relation within a single superconducting circuit.


Here, we introduce a hybrid insulator- and semiconductor-based circuit element in which the parity of the Josephson harmonics can be tuned, enabling orders-of-magnitude adjustments of the relative strengths of even and odd components of the Josephson potential. 
The focus of this work is the realization and \emph{in situ} control of higher Josephson harmonics--and in particular their parity--within a single compact superconducting circuit element. 
We refer to this element as the Harmonic Parity Qubit (HPQ).

Our device combines two SIS tunnel junctions in one arm of a superconducting quantum interference device (SQUID), in parallel with a gate-tunable S-Sm-S nanowire junction~\cite{doh2005tunable} in the other, as shown schematically in Fig.~\ref{Fig1}b. 
This configuration realizes a compact Josephson element in which SIS and S-Sm-S junctions share the same phase degree of freedom and contribute coherently to a single composite potential. 
The harmonic control arises from the interference between the fixed harmonic content of the SIS-SIS arm and the gate-tunable harmonics of the S-Sm-S junction.

Specifically, the SIS-SIS arm provides a stable higher-harmonic backbone~\cite{bozkurt2023double,Jacobsen_SISSIS_2025, shagalov2025, banszerus2024voltage}, while the nanowire contributes a nonsinusoidal energy-phase relation with continuously gate-tunable transmission channels and thereby tunable prefactors of the $\cos(k\phi)$ terms in the Josephson potential.
At half-flux quantum frustration, odd harmonics from the two arms cancel each other, enabling us to sweep between three distinct regimes: a conventional $\cos(\phi)$-dominated potential (Fig.~\ref{Fig1}a, left panel), a mixed-harmonic regime where $\cos(\phi)$ and $\cos(2\phi)$ are comparable, and an even-harmonic regime in which $\cos(\phi)$ and other odd components are significantly suppressed (Fig.~\ref{Fig1}a, middle panel). 
By fitting multi-transition spectroscopy across 85 gate settings, we directly reconstruct the full harmonic landscape and demonstrate suppression of the fundamental harmonic term by more than two orders of magnitude relative to the second harmonic, yielding a robust $\pi$-periodic potential.

\section*{Results}
In Fig.~\ref{Fig1}b, we show a lumped-element circuit diagram of the HPQ device, which we use to demonstrate control of the Josephson harmonic content. After eliminating a high-energy mode, the circuit can be approximated by the Hamiltonian
\begin{equation}
    \label{eq:Ham}
    \hat{H}_\text{\qubitacr} = 4 E_{C} {(\hat{n}-n_g)}^2 +\hat{U}_{\sissis}(\hat{\phi}) + \hat{U}_{\sns}(\hat{\phi};V_g,\varphi_\text{e}),
\end{equation} 
where $E_C$ is the charging energy of the island, $\hat{n}$ is the charge operator, $n_g$ is the gate charge of the island, $\hat{\phi}$ is the phase operator, $\varphi_\mathrm{e}=2\pi\Phi_\mathrm{e}/\Phi_0$ is the reduced flux ($\Phi_0$ is the fundamental flux quantum), $\Phi_\mathrm{e}$ is the external flux threading the loop, and $V_g$ is the gate voltage that tunes the harmonic content of the S-Sm-S junction. Furthermore, $U_{\sissis}$ refers to the potential energy of the arm with the two SIS junctions, and $U_{\sns}$ is the potential energy of the single S-Sm-S junction.

The fixed higher harmonics contribution from the SIS-SIS element can be seen from their total potential energy, which, after eliminating the high-energy mode associated with the middle superconducting island, is approximately given by~\cite{bozkurt2023double, Jacobsen_SISSIS_2025},
\begin{align}
    U_{\sissis}(\phi) &= -E_{J\Sigma}\sqrt{1-\lambda\sin^2\left(\phi/2 \right)} \nonumber \\
    & = \sum_{k=0}^\infty {u}_k \cos(k\phi), \label{eq:sissis_harmonics}
\end{align}
where $E_{J\Sigma} = E_{J,1}+E_{J,2}$ is the sum of the Josephson energies of the junctions  $E_{J,1}$ and $E_{J,2}$, and $\lambda = 4E_{J,1}E_{J,2}/(E_{J,1}+E_{J,2})^2$ is the asymmetry of the two junctions.
The coefficients ${u}_k$ are the Fourier amplitudes of $U_{\sissis}$ (see  Supplementary Section~I), and they set the harmonic content of the potential.
The higher harmonics ($k>2$) can reach ${u}_2/{u}_1 \approx 20\%$ for $\lambda = 1$~\cite{shagalov2025}.

The gate-tunable hybrid element has a structurally similar potential given by~\cite{beenakker1991universal, larsen2015semiconductor, kringhoj2018anharmonicity, david_paper}
\begin{align}
    U_{\sns}(\phi;V_g,\varphi_\text{e})&=-\Delta\sum_i\sqrt{1-T_i(V_g)\sin^2\left((\phi-\varphi_\text{e})/2 \right)} \nonumber\\
    & = \sum_{k=0}^\infty {v}_k(V_g)\cos(k[\phi - \varphi_\text{e}]),
\end{align}
where $\Delta$ is the superconducting gap, $T_i(V_g)$ is the gate-tunable transmission of the $i$'th channel in the junction, and the sum runs over all conduction channels. 
The coefficients ${v}_k(V_g)$ are the Fourier amplitudes of $U_{\sns}$ and depend on the gate voltage through the channel transmissions $\{T_i(V_g)\}$ (see Supplementary Section~I). 
The higher harmonics in the hybrid element arise due to higher-order coherent tunneling processes via Andreev bound states in the semiconducting region~\cite{beenakker1991universal}.

The ability to control the harmonic parity in the \qubitacr{} can be seen by considering the total  potential and its Fourier expansion,
\begin{align}
    U_\text{\qubitacr} & = U_{\sissis} + U_{\sns}\\
    & = \sum_{k=0}^\infty {c}_k(V_g, \varphi_\text{e}) \cos(k\phi) +  {s}_k(V_g, \varphi_\text{e})\sin (k\phi),\nonumber
\end{align}
where the prefactors ${c}_k$ and ${s}_k$ are given in Supplementary Section~I.
Crucially, at $\varphi_\text{e} = \pi$ these prefactors reduce to
\begin{equation}
\begin{split}
    {c}_k(V_g,\pi) &= {u}_k + (-1)^k {v}_k(V_g),\\ 
    {s}_k(V_g,\pi) &= 0.
\end{split}
\end{equation}

Thus, at this flux value, the coefficients of the sine components vanish, and the cosine prefactors reduce to the sum or difference of the cosine coefficients of the two arms, depending on their parity. The leading Fourier coefficients $u_k$ and $v_k$ have the same sign, as both the SIS-SIS and S-Sm-S contributions arise from Josephson potentials with the same qualitative shape, namely even functions of $\phi$ with minima at $\phi=0$. As a result, at $\varphi_e=\pi$, odd-$k$ harmonics tend to interfere destructively, while even-$k$ harmonics interfere constructively. In particular, when the harmonic coefficients are matched, ${u}_k = {v}_k(V_g)$, the odd-$k$ terms cancel while the even-$k$ terms add, so that only even-$k$ harmonics remain nonzero. Thus, the joint flux and gate control in the \qubitacr{} turns harmonic parity into a programmable property at $\varphi_\text{e} = \pi$, enabling selective realization of odd- or even-dominated Josephson potential in a single, compact element as shown schematically in Fig.~\ref{Fig1}b.

\begin{figure}[t!]
\includegraphics{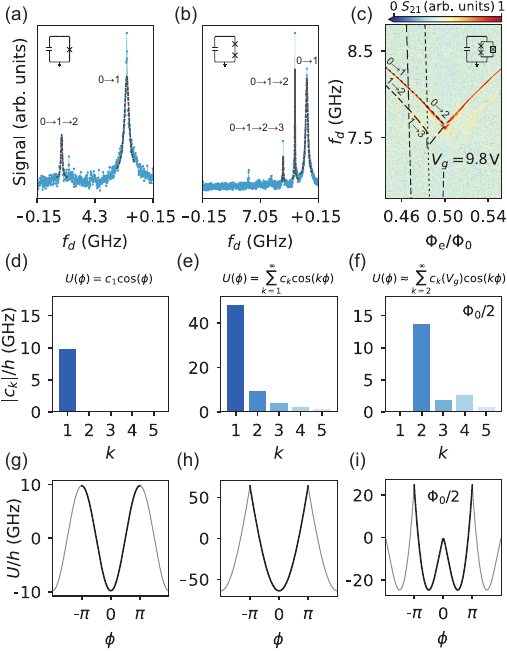}
\caption{\label{Fig2} Circuit-engineered Josephson harmonics extracted from qubit spectroscopy. Two-tone spectroscopy of (a) a transmon, (b) a double-junction transmon, (c) the Harmonic parity qubit at $V_g = \SI{9.8}{V}$ versus external flux $\Phi_\text{e}$ (see main text for details). 
In (a),(b) the signal trace corresponds to an average of qubit spectroscopy measurements over different drive powers. 
Dashed lines are fits to the identified transition frequencies used to extract the Hamiltonian parameters. 
(d)-(f) Harmonics of the potential in absolute value using parameters extracted from the fit. Each potential is decomposed into harmonics via a Fourier series $U(\phi)=\sum_{k=1}^{\infty} {c}_k \cos(k\phi)$. The transmon transitions are well described by a $\cos(\phi)$ term, while the double-junction transmon shows appreciable higher harmonics. For the \qubitacr{} at half flux quantum, the first harmonic is strongly suppressed, and $\cos(2\phi)$ dominates. (g)-(i) show the potential calculated using fit parameters. The transmon potential is sinusoidal, while the double-junction transmon shows a more parabolic well. The \qubitacr{} potential at half flux quantum shows a double well dominated by a $\cos(2\phi)$-term.}
\end{figure}

To demonstrate the harmonic parity control in the HPQ, we fabricated a chip that integrates three nominally identical qubit islands implemented with increasing Josephson complexity: a conventional transmon, a double-junction SIS-SIS transmon, and the gate-tunable \qubitacr.
All devices share the same capacitive island geometry (Fig.~\ref{Fig1}c), while the HPQ incorporates an InAs/Al hybrid nanowire junction in one arm of the SQUID loop, with channel transmissions tunable by an external gate voltage $V_g$.
Figures~\ref{Fig1}c-e show scanning electron micrographs of the \qubitacr{} device.

First, we consider a reference qubit in which the first harmonic dominates the potential, namely the single-junction transmon. Two-tone spectroscopy (Fig.~\ref{Fig2}a) reveals the transitions $f_{01}$ and $f_{02}/2$. In fitting this device, we assume a single-harmonic Josephson potential, since only these two transitions are experimentally resolved and no higher-order transitions are observed. Using the standard transmon Hamiltonian~\cite{koch2007charge}, we extract $E_C/h = \SI{280}{MHz}$ and $E_J/h = \SI{9.8}{GHz}$. The corresponding energy-phase relation is well described by a single $\cos(\phi)$ term (Fig.~\ref{Fig2}d), providing a reference potential and geometric device parameters (Fig.~\ref{Fig2}g) against which to compare our more complex elements.

Adding a second SIS junction in series dramatically reshapes the spectrum. 
In the double-junction SIS-SIS transmon (Fig.~\ref{Fig2}b), we observe pronounced changes in anharmonicity and identify the transitions $f_{01}$, $f_{02}/2$, and $f_{03}/3$. 
Fitting the data with the double-junction Hamiltonian [Eq.~\eqref{eq:sissis_harmonics}] yields $E_{J,1}/h = E_{J,2}/h = \SI{59.96}{GHz}$ and a junction self-capacitance energy $E_{C_J}/h = \SI{583}{MHz}$ associated with the high-energy mode (see Supplementary Section~I). 
Fourier decomposition of the extracted energy-phase relation shows appreciable higher harmonics, with a second harmonic whose magnitude relative to the first one is $|{u}_2/{u}_1| = 0.19$ (Fig.~\ref{Fig2}e), consistent with theory~\cite{shagalov2025}.
The harmonic content also exhibits smaller but detectable third- and fourth-order components whose amplitudes decay monotonically with increasing $k$.
The resulting nonsinusoidal qubit potential produces a more quadratic well (Fig.~\ref{Fig2}h).

We then combine this fixed harmonic backbone with the gate-tunable nanowire junction to realize the full \qubitacr{} element (inset of Fig.~\ref{Fig2}c).
Flux-dependent spectroscopy of this device (Fig.~\ref{Fig2}c) shows a rich pattern of transitions $f_{01}$, $f_{12}$, $f_{02}$, and $f_{13}$, which we fit using the full circuit Hamiltonian (see Supplementary Section~III for details), extracting $E_{J,1}/h = E_{J,2}/h = \SI{55.03}{GHz}$, $E_{C_J}/h = \SI{675}{MHz}$, and $\Delta/h = \SI{40.06}{GHz}$, consistent with materials and fabrication parameters (see Supplementary Section~II). 
At half-flux quantum, the harmonic landscape is qualitatively different from the double-junction case: the reconstructed spectrum at $\Phi_\text{e} = \Phi_0/2$ is dominated by the $\cos(2\phi)$ term, with the fundamental $\cos(\phi)$ component suppressed and the next odd harmonic $\cos(3\phi)$ reduced relative to $\cos(4\phi)$, consistent with an even-parity-dominated Josephson potential (Fig.~\ref{Fig2}f,i).

As the overall harmonic content of the HPQ is set by the balance between the two arms of the SQUID loop, the gate voltage $V_g$ provides the crucial knob for controlling this balance. 
We therefore perform flux-dependent spectroscopy at 85 gate voltage settings, stepping $V_g$ from $-7~\text{V}$ to $9.8~\text{V}$ in increments of $0.2~\text{V}$. 
For each gate value, we perform resonator spectroscopy to determine the readout frequency and then apply a continuous microwave drive tone through the gate line to perform qubit spectroscopy.
We then fit the observed qubit transitions with the full circuit Hamiltonian, using a single set of global device parameters $(E_{J,1},E_{J,2},E_{C_J},E_C,\Delta)$ and allowing only the set of transmission coefficients $T_i(V_g)$ of the nanowire channels to vary.
This fitting protocol, detailed in Supplementary Section~III, yields the potential at each gate voltage, from which we extract the harmonic coefficients ${c}_k(V_g,\varphi_\text{e})$ and ${s}_k(V_g,\varphi_\text{e})$.

\begin{figure*}[t!]
\includegraphics{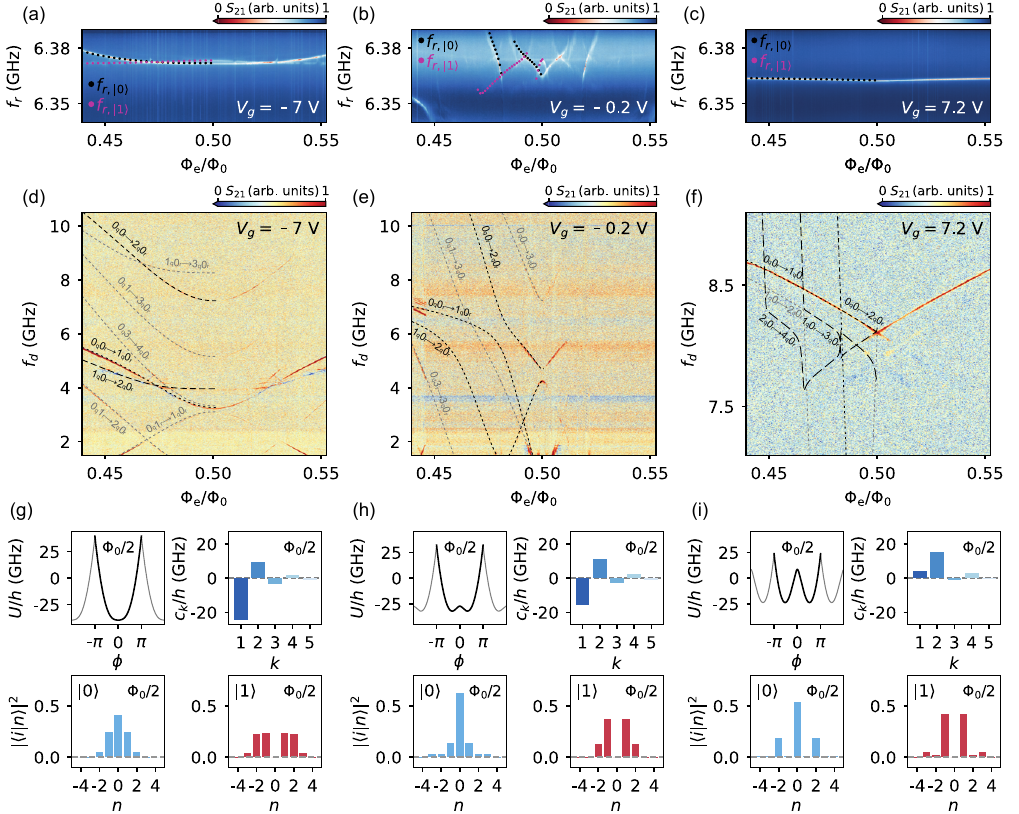}
\caption{\label{Fig3} Gate evolution of the HPQ spectrum, potential and wavefunctions. Spectroscopy of the \qubitacr{} as a function of external flux $\Phi_\text{e}$ for three different gate voltages: (a) $V_g = \SI{-7}{V}$, (b) $V_g = \SI{-0.2}{V}$, and (c) $V_g = \SI{7.2}{V}$. Solid points denote simulation of the resonator frequency when the qubit is in the ground (black) and when the qubit is in the excited state (pink), using parameters from the qubit spectroscopy fit. (d)-(f) Two-tone qubit spectroscopy as a function of flux at the corresponding $V_g$. Black dashed lines indicate transitions that have been used for the fit, while gray dashed lines indicate other transitions we have identified but not used for the fit, some of which arise from the resonator being thermally excited. (g)-(i) Potential (top-left), harmonics (top-right), and charge-number wavefunction amplitudes for ground $\ket{0}$ (bottom-left) and first excited $\ket{1}$ (bottom-right) states calculated at half flux quantum using parameters extracted from the fit, showing the transition to qubit states identified by opposite Cooper pair parity.}
\end{figure*}

We first highlight three representative gate voltages that illustrate how the \qubitacr{} evolves from a conventional, $2\pi$-periodic regime into an even-harmonic-dominated regime. 
At the most negative gate, $V_g=\SI{-7}{V}$, the fitted transmissions of the nanowire junction are $T_i=\{0.68,0.47,0.46\}$, and the spectrum in Fig.~\ref{Fig3}a,d exhibits a typical flux-qubit-like dependence.
The reconstructed potential at $\Phi_\text{e}=\Phi_0/2$ (Fig.~\ref{Fig3}g) is $2\pi$-periodic with a dominant $\cos\phi$ component and visible cusps near $\phi=\pi$ arising from smaller higher-order harmonics, resulting in a “quarton-like” single-well landscape \cite{yan2020engineering}.
The wavefunctions in charge representation contain even and odd charge states (Fig.~\ref{Fig3}g bottom) for both the ground $\ket{0_\text{q}}$ and excited $\ket{1_\text{q}}$ state, similar to the transmon qubit (assuming $n_g=0$ for clarity).

\begin{figure}[t!]
\includegraphics{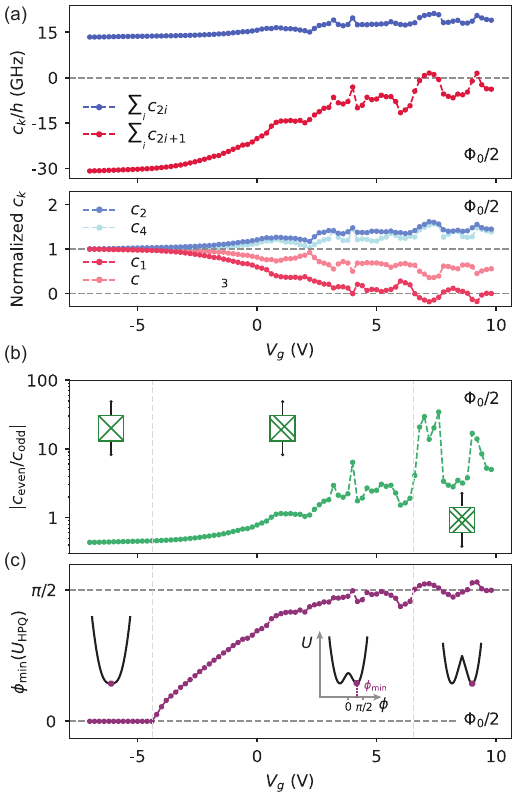}
\caption{Gate-tunable control of Josephson harmonic parity. (a) Harmonic content versus gate voltage. Top: The sum of even (blue) and odd (red) coefficients $c_k$ as a function of gate voltage. The odd components $c_\mathrm{odd}$ are progressively suppressed with increasing $V_g$ and change sign, implying $c_\mathrm{odd}=0$ when it crosses zero, while $c_\mathrm{even}$ remains large and varies slowly. Bottom: Individual even (blue) and odd (red) coefficients, each normalized to its value at $V_g=\SI{-7}{V}$. All odd components decrease, whereas all the even components increase. In particular, $c_1$ crosses zero at multiple gate points. All coefficients are extracted by fitting the qubit spectrum at each gate. 
(b) Harmonic parity ratio $|c_\mathrm{even}/c_\mathrm{odd}|$ versus gate voltage. Tunability of $|{c}_\text{even}/{c}_\text{odd}|$ spans two orders of magnitude across the measured gate range and formally diverges where $|{c}_\text{odd}|$ crosses zero. 
(c) Phase point $\phi_\mathrm{min}$ where the qubit potential is minimum for each $V_g$. Dashed vertical lines demarcate three regimes based on $\phi_\mathrm{min}$: a $\cos(\phi)$-dominated regime, a comparable $\cos(\phi)$ and $\cos(2\phi)$ regime, and an even-harmonic dominated regime consistent with a $\pi$-periodic potential. Insets show the potential at representative gate voltages (left to right) $V_g=\SI{-6.0}{V}$, $V_g=\SI{2.4}{V}$, $V_g=\SI{9.2}{V}$.\label{Fig4}}
\end{figure}

Increasing the gate to $V_g=-0.2~\text{V}$ substantially modifies the nanowire channels, with the fit yielding $T_i=\{0.94,0.58,0.58\}$.
The resulting fit and transitions are shown in Fig.~\ref{Fig3}b,e.
In this regime, the qubit spectrum (Fig.~\ref{Fig3}e) develops pronounced flux dispersion, characteristic of flux-qubit behavior \cite{van2000quantum,chiorescu2003coherent}.
At half flux quantum, $f_{01}$ drops below \SI{1.5}{GHz}, while the charge matrix element for the $\ket{0_\text{q}}\rightarrow \ket{2_\text{q}}$ transition nearly vanishes, and an avoided crossing between $f_{01}$ and $f_{02}$ remains, set by the charge matrix element between the first $\ket{1_\text{q}}$ and second $\ket{2_\text{q}}$ excited states, i.e. $\bra{1_\text{q}}n\ket{2_\text{q}}$. 
The extracted harmonics at $\Phi_\text{e}=\Phi_0/2$ show that destructive interference between the $\cos(\phi)$ terms of the two SQUID arms suppresses the net first harmonic and enhances $\cos(2\phi)$, resulting in $|{c}_2/{c}_1|=0.71$ and generating a shallow double-well potential similar to a flux qubit with a small barrier centered at $\phi=0$ (Fig.~\ref{Fig3}h).
The charge wavefunctions (Fig.~\ref{Fig3}h bottom) in this mixed-harmonic regime reflect a slight parity bias: the ground state moderately supports even-$k$ components, while the first excited state supports slightly more odd-$k$ components. An additional dataset in this regime is shown in Supplementary Section~V, where the qubit frequency drops to $f_{01}=\SI{191}{MHz}$ at half flux quantum.

Increasing the gate further to $V_g=\SI{7.2}{V}$ drives the device into a regime where even harmonics dominate the potential energy. 
Here, the fitted transmissions increase to $T_i=\{0.98,0.98,0.75,0.54\}$. The qubit spectrum (Fig. ~\ref{Fig3}c,f) exhibits the characteristic level structure of a deep double well: the avoided crossing between $f_{01}$ and $f_{02}$ becomes smaller than our experimental resolution, consistent with a strongly suppressed transition matrix element between the first and second states due to opposite Cooper-pair parity of the two states. 
The corresponding potential and harmonic decomposition at $\Phi_e=\Phi_0/2$ (Fig.~\ref{Fig3}i) reveal that the $\cos(\phi)$ coefficient is small while $\cos(2\phi)$ remains large ($|{c}_2/{c}_1|=3.47$) and higher even components are still appreciable. Thus, the double-well landscape is governed by the significant contributions of even harmonics. The dominant even harmonics separate the charge states of the ground $\ket{0_\text{q}}$ and excited $\ket{1_\text{q}}$ state cleanly by parity (Fig.~\ref{Fig3}i bottom), such that $\ket{0_\text{q}}$ and $\ket{1_\text{q}}$ has support on even and odd charge states respectively, while the opposite-parity components are strongly suppressed.
As $V_g$ is swept between these three regimes, the fitted ${c}_1(V_g,\pi)$ changes sign, implying that there exists an intermediate gate voltage at which the first harmonic crosses zero and the $\cos(\phi)$ term is canceled exactly.

Figure~\ref{Fig4} summarizes the central result of this work: continuous tuning from an odd-harmonic-dominated regime to an even-harmonic-dominated regime, with a well-defined gate point where the odd contributions cancel.
To obtain a global picture of how gate control programs the HPQ, we extract the harmonic coefficients ${c}_k(V_g,\varphi_\text{e})$ at $\Phi_\text{e}=\Phi_0/2$ for all 85 measured gate voltages. 
Figure~\ref{Fig4}a (top) shows the sum of the even-harmonic weights ${c}_\text{even}=\sum_i {c}_{2i}$ (blue) and odd-harmonic weights ${c}_\text{odd}=\sum_i {c}_{2i+1}$ (red) as a function of $V_g$. 
As the gate is swept from $-7~\mathrm{V}$ to $9.8~\mathrm{V}$, the total odd contribution collapses from $|{c}_\text{odd}|/h= \SI{30.9}{GHz}$ to $\SI{0.6}{GHz}$ at its minimum, while the total even contribution remains large and comparatively stable in the range $|{c}_\text{even}|/h= 13.4{-}21.2~\mathrm{GHz}$. 
Thus, the gate-tunability of the HPQ can suppress odd components at $\Phi_\text{e} = \Phi_0/2$ while preserving sizeable even-harmonic terms of order $15{-}\SI{20}{GHz}$. 
The maximum second-harmonic amplitude reaches $c_2^{\mathrm{max}}/h = \SI{15.3}{GHz}$ (at $V_g = \SI{7.2}{V}$, Fig.~\ref{Fig3}c), which corresponds to $c_2^{\mathrm{max}}/E_C \approx 54$, placing the device well within the transmon regime even in the even-harmonic-dominated limit. 
This behavior arises because the potential terms in the two SQUID arms interfere destructively for odd harmonics and constructively for even harmonics.

To resolve how individual harmonics participate in this evolution, Fig.~\ref{Fig4}a (bottom) plots ${c}_1,{c}_2,{c}_3$ and ${c}_4$ as a function of $V_g$, each normalized to their respective value at $V_g=-7~\mathrm{V}$. 
The even terms ${c}_2$ and ${c}_4$ (blue) increase in magnitude with gate voltage, whereas the odd terms ${c}_1$ and ${c}_3$ (red) decrease and eventually change sign. 
In particular, ${c}_1(V_g)$ crosses through zero while higher even harmonics remain finite.
Although an equivalent suppression of ${c}_1$ could be achieved with a design with a single SIS junction in parallel with an S-Sm-S junction, the double-SIS arm of the \qubitacr{} contributes its own higher-order odd components, enabling suppression of \textit{all} odd harmonics. 
A two S-Sm-S junction design \cite{larsen2020parity} can, in principle, achieve similar multi-odd harmonics cancellations, but requires coordinating two gates on two junctions that may be unstable over time \cite{david_paper}. Here, a single gate tunes the S-Sm-S harmonic coefficients to match those of the fixed SIS-SIS backbone, simplifying operation and improving stability.

A compact way to summarize the change in overall harmonic parity is to consider the ratio $|{c}_\text{even}/{c}_\text{odd}|$, shown in Fig.~\ref{Fig4}b. 
This even-to-odd metric is tunable over nearly two orders of magnitude across the accessible gate range and formally diverges where $|{c}_\text{odd}|$ passes through zero. 
To connect the harmonic parity to the potential landscape of the qubit, Fig.~\ref{Fig4}c shows the phase $\phi_{\min}(U_\text{\qubitacr})$ at which the potential $U_\text{\qubitacr}(\phi)$ reaches its minimum for each $V_g$. 
We identify three distinct regimes. In an odd-parity dominated regime, $\phi_{\min}= 0$ and the potential is $2\pi$-periodic (quarton qubit-like). In the intermediate mixed regime, comparable odd and even harmonic content shifts the minimum away from zero and generates a shallow double well (flux qubit-like). In an even-dominated regime, odd harmonics are strongly suppressed, while even harmonics remain large, shifting the minima to $\phi_{\min}\approx \pm\pi/2$ and yielding an effectively $\pi$-periodic potential governed by even harmonics.
This classification via $\phi_{\min}(U_\text{\qubitacr})$ makes it explicit that a single gate parameter suffices to steer the \qubitacr{} between $2\pi$- and $\pi$-periodic behavior.

In summary, we have realized a hybrid Josephson element in which a double SIS junction and a gate-tunable S-Sm-S junction cooperate to make harmonic parity a programmable resource. 
By combining detailed multi-transition spectroscopy with a global circuit model across 85 gate settings, we reconstruct the gate- and flux-dependent harmonic landscape and show that the total even contribution remains at the 15-20 GHz scale while the odd contribution can be exactly canceled. 
This control lets us tune the relative weight of multi-Cooper-pair tunneling processes, enabling a new axis of circuit design: parity selection, tailored energy barriers, and programmable nonlinearity on demand. 
Across the measured gate range, the ratio $|{c}_\mathrm{even}/{c}_\mathrm{odd}|$ spans from $0.4$ to $\approx 34.5$, including gate points where ${c}_\mathrm{odd}$ changes sign and cancels. 
In the even-harmonic-dominated regime, the Josephson potential forms a deep double well dominated by $\cos(2\phi)$ and higher even harmonics, with minima near $\phi \approx \pm\pi/2$. 
Importantly, this even-dominated regime is reached while remaining in the transmon limit, showing that a strongly $\cos(2\phi)$-dominated potential remains compatible with transmon operation.
More broadly, our results demonstrate that the harmonic content, and in particular the parity structure, of the Josephson potential can be engineered \emph{in situ} within a compact element, opening a new pathway to intrinsically protected qubits, tailor-made couplers~\cite{maiani2022entangling}, and Hamiltonian engineering schemes that exploit multi-Cooper-pair tunneling in complex superconducting quantum circuits.

\section*{Data availability}

The data that support the finding of this work is available from the Zenodo repository at
https://doi.org/10.5281/zenodo.21396973~\cite{feldstein_bofill_2026_21396973}.

\section*{Acknowledgments}

We gratefully acknowledge Peter Krogstrup for nanowire growth. We also gratefully acknowledge useful discussions with Christian Andersen and Nataliia Zhurbina. Finally, we gratefully acknowledge Lena Jacobsen for program management support.

\section*{Funding Statement}

This research was supported by the Novo Nordisk Foundation (grant no. NNF22SA0081175), the NNF Quantum Computing Programme (NQCP), Villum Foundation through a Villum Young Investigator grant (grant no. 37467), the Innovation Fund Denmark (grant no. 2081-00013B, DanQ), the U.S. Army Research Office (grant no. W911NF-22-1-0042, NHyDTech), by the European Union through an ERC Starting Grant, (grant no. 101077479, NovADePro), and by the Carlsberg Foundation (grant no. CF21-0343). 
Any opinions, findings, conclusions or recommendations expressed in this material are those of the author(s) and do not necessarily reflect the views of Army Research Office or the US Government. 
Views and opinions expressed are those of the author(s) only and do not necessarily reflect those of the European Union or the European Research Council. Neither the European Union nor the granting authority can be held responsible for them. 

\section*{Author Contributions Statement}

D.F.B., Z.S., S.K., K.F. and M.K. designed the experiment. S.K. and M.K. supervised the project, with additional input from A.G., J.H. and A.K. D.F.B. designed, fabricated and measured the devices. L.U.J., K.S. and S.K. developed the theoretical model for the double-junction circuit. Z.S. contributed to the microwave measurements. Z.S., C.W. and S.S. contributed to the nanowire-junction nanofabrication. D.F.B. analyzed the data. D.F.B., S.K. and M.K. wrote the manuscript with input from all co-authors. All authors discussed the results and contributed to the manuscript.

\section*{Competing Interests Statement}
The authors declare no competing interests.

\appendix 

\section{Circuit Hamiltonian}\label{Sec:Hamiltonian}

The SIS Josephson junction provides the transmon qubit with a periodic potential of the form $U_\text{SIS}=-E_J \cos(\phi)$ \cite{koch2007charge}, where $E_J$ is the Josephson energy and $\phi$ is the superconducting phase difference across the junction. 

A potential with higher harmonics arises from placing two SIS tunnel junctions in series \cite{bozkurt2023double, shagalov2025}. 
In this way, the total phase is distributed across the two junctions, and the classical potential energy ($U_\text{SIS+SIS}$) matches that of an S-Sm-S junction with finite transparency:

\begin{equation}
    U_\text{SIS+SIS}(\phi) = - E_{J\Sigma} \sqrt{1 - \lambda \sin^2{(\phi/2)}},
\end{equation}

where $E_{J\Sigma} = E_{J,1}+E_{J,2}$ with $E_{J,1},E_{J,2}$ the Josephson energies of junction 1 and 2, and where $\lambda = 4E_{J,1}E_{J,2}/(E_{J,1}+E_{J,2})^2$ is the effective transmission of the element.
When such a Josephson element is introduced in a transmon architecture, the anharmonicity of the qubit decreases to $1/4$ ($\lambda=1$) of its single junction equivalent \cite{shagalov2025}, just like in the S-Sm-S based transmon when the ABS transmission is unity ($T=1$) \cite{kringhoj2018anharmonicity}.

Another consequence of placing two SIS junctions in series is the emergence of a second mode in the device whose charging energy is set by the intrinsic Josephson junction capacitances $C_J$. Although we seek a model for the qubit mode only, we have to account for the presence of this internal mode as it renormalizes the qubit energy spectrum \cite{Jacobsen_SISSIS_2025}. 
The effect of the internal mode on the qubit spectrum can be accounted for in the Born-Oppenheimer (BO) approximation \cite{Jacobsen_SISSIS_2025}, which integrates out the contribution of the internal mode on energy levels lower than the first internal mode excitation.
Thus, the internal mode adds a correction term $U_\text{BO}$ to the double junction potential:

\begin{equation}
    U_\text{\sissis} = U_\text{SIS+SIS}+U_\text{BO}
\end{equation}

Assuming both junctions have the same intrinsic capacitance $C_J$, the Born-Oppenheimer potential term reads:

\begin{equation}\label{eq:U_BO_A}
    U_\text{BO}(\phi) = E_{J\Sigma}\sqrt{\frac{E_{C_J}}{E_{J\Sigma}}\sqrt{1-\lambda \sin^2{({\phi}/2)}}},
\end{equation}

where we defined the charging energy of the junctions $E_{C_J}=e^2/2C_J$.
There are two key aspects to consider. 
First, this correction is valid when the internal mode energy is higher than the qubit energy levels of interest. 
The frequency of the lowest internal mode transition can be approximated by $f_\text{int} \approx\sqrt{4E_{C_J}E_{J\Sigma}}/h$. 
Thus, for a transmon-like qubit mode, the approximation holds when $E_{C_J}> E_C$, where $E_C = e^2/(2C_q+C_J)$ is the qubit charging energy and $C_q$ is the qubit island capacitance.
Second, performing the harmonic approximation of the internal mode ground state to arrive at Eq.~\eqref{eq:U_BO_A} requires $E_{J\Sigma}/E_{C_J}\gg1$. 
Therefore, to minimize the contribution of the internal mode and to be able to operate in the Born-Oppenheimer regime, we need to design the Josephson junctions such that $E_{J\Sigma}/E_{C}> E_{J\Sigma}/E_{C_J}\gg1$, which in turn requires that $C_q > C_J$.

The total Hamiltonian of the system is then given by

\begin{equation}
    \label{eq:HamT}
    \hat{H} = 4 E_{C} (\hat{n}-n_g)^2 +{U}_\text{SIS+SIS} + {U}_\text{BO}+{U}_\text{\sns}(V_g,\varphi_\text{e})
\end{equation}

\begin{figure*}[t!]
\centering
\includegraphics{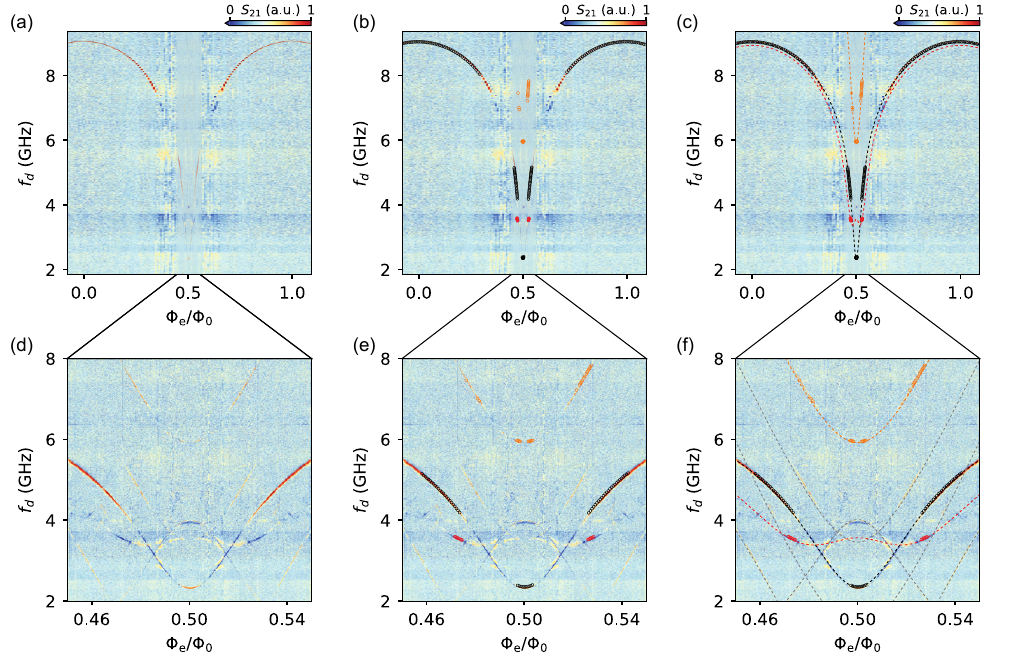}
\caption{Fitting routine. (a)/(d) Wide-range/narrow-range two-tone spectroscopy as a function of applied flux. (b),(e) Same data with extracted transition frequencies overlaid: $f_{01}$ (black markers), $f_{12}$ (red markers), and $f_{02}$ (orange markers). These are the transitions used in the fit. (c),(f) Global fit of the qubit Hamiltonian to the extracted points. Black/orange/red dashed curves show the fitted transitions, while gray dashed lines show transitions we identified that were not used for the fit. \label{Fig3_app}}
\end{figure*}

We can expand the total potential as a Fourier series with cosine ($c_k$) and sine ($s_k$) Fourier components
\begin{equation}
\begin{split}
    {c}_k(V_g,\varphi_\mathrm{e}) &= u_k + \cos(k\varphi_\mathrm{e})v_k(V_g),\\ 
    {s}_k(V_g,\varphi_\mathrm{e}) &= \sin(k\varphi_\mathrm{e})v_k(V_g).
\end{split}
\end{equation}
At half flux quantum, the Fourier coefficients reduce to:

\begin{equation}
\begin{split}
    {c}_k(V_g,\pi) &= [{u}_k + (-1)^k {v}_k(V_g)]\\ 
    {s}_k(V_g,\pi) &= 0.
\end{split}
\end{equation}

where the sine components vanish because the total potential $U(\phi)$ is even in $\phi$. Here, $u_k$ and $v_k$ are the $k$-th cosine Fourier amplitudes of the SIS-SIS branch (including the BO correction) and of the S-Sm-S branch, respectively:

\begin{equation}
    {u}_k = \frac{1}{2\pi}\int_{-\pi}^{\pi} \big[U_\text{SIS+SIS}({\phi'}) + U_\text{BO}({\phi'})\big] \cos(k {\phi'}) d{\phi'}
\end{equation}

and

\begin{equation}
    {v}_k = \frac{1}{2\pi}\int_{-\pi}^{\pi} U_{\sns} ({\phi'}) \cos(k {\phi'}) d{\phi'}.
\end{equation}

\section{Fabrication}\label{Sec:Fab}

Our devices are fabricated on a $\SI{70}{nm}$ thick NbTiN superconducting layer sputtered onto a high-resistivity silicon substrate. The device is patterned via dry reactive-ion etching using an $\mathrm{SF}_6/\mathrm{O}_2$ plasma. A small region containing the gate tip is subsequently etched to create a vacuum gap between the gate and the nanowire. The SIS junction arms are evaporated in a Plassys system via double-angle evaporation with in situ oxidation. An InAs nanowire with an epitaxial Al shell grown on all facets is then deterministically placed on the contact pads using a micromanipulator, such that the nanowire is suspended between the two contacts~\cite{david_paper,lu2025andreev}. The S-Sm-S Josephson junction is formed by selectively etching away a $\sim\SI{200}{nm}$-long segment of the $\sim\SI{30}{nm}$-thick Al shell. Finally, the two ends of the nanowire and the SIS junction arms are connected to the control layer using Al patches, with a previous Argon milling step to ensure good electrical contact.

Table \ref{tab:qubit_params} shows an overview of the simulated/extracted parameters from all devices on the chip. 

\begin{table}[h!]
    \centering
    \begin{tabular}{||c c c c c||} 
     \hline
     Qubit & $E_C/h (\text{MHz})$ & $C_{c} (\text{fF})$ & $C_d (\text{fF})$ & $A_\text{JJ} (\mathrm{\mu m^2})$ \\ [0.5ex] 
     \hline\hline
     Transmon  & 280 & 4.9 & N.A. & 0.12 \\ 
     \hline
     D.J. Transmon A  & 280 & 4.9 & 0.33 & 0.61 \\
     \hline
     D.J. Transmon B  & 280 & 4.9 & N.A. & 0.18 \\     
     \hline
     \qubitacr{}  & 280 & 4.9 & 0.33 & 0.59 \\
     \hline
    \end{tabular}
    \caption{Summary of qubit parameters showing 1) Qubit type, 2) Extracted qubit charging energy, 3) Simulated qubit-resonator coupling capacitance, 4) Simulated qubit-gate coupling capacitance, 5) Area of the SIS Josephson junctions extracted from scanning electron microscope.}
    \label{tab:qubit_params}
\end{table}

\begin{figure*}[t!]
\centering
\includegraphics{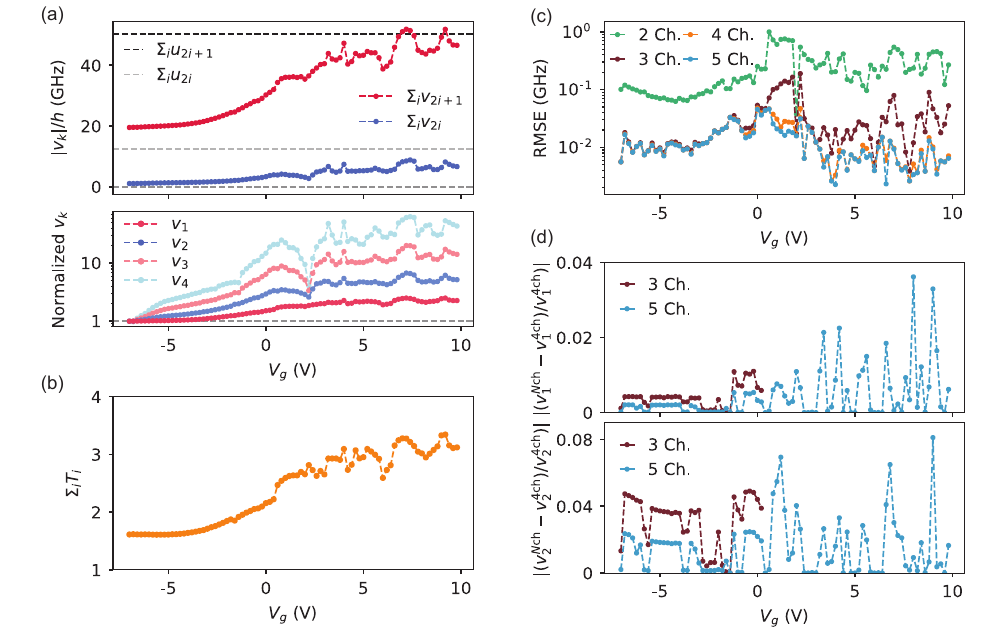}
\caption{Gate dependence and fit robustness of the extracted S-Sm-S harmonics. (a) Top: The sum of even (blue) and odd (red) S-Sm-S Fourier coefficients $v_k$ as a function of gate voltage. The fixed contributions from the double-SIS branch are shown as black (sum of odd) and gray (sum of even) dashed lines. Bottom: Individual even (blue) and odd (red) coefficients, each normalized to its value at $V_g=\SI{-7}{V}$. All coefficients are extracted by fitting the qubit spectrum at each gate. (b) The sum of even (odd) coefficients, shown in blue (red), as a function of gate voltage. (c) Sum of the fitted transmission channels at each $V_g$. Datasets with $V_g<\SI{0.6}{V}$ are fitted with three channels; while datasets with $V_g\geq\SI{0.6}{V}$ are fitted with four channels. (c) Root mean square error (RMSE) of the fitted transition frequencies of each flux scan at every gate voltage, for models including two to five conduction channels. The RMSE quantifies the average deviation between experimental and best-fit transition frequencies. The fit quality improves significantly when increasing the number of channels from two to three, while higher-channel models (four and five) further improve the fit only for $V_g>0$. (d) Relative difference between the S-Sm-S harmonics coefficients $v_1^{N\text{ch}}$ and $v_2^{N\text{ch}}$ of the multi-channel models ${N\text{ch}}=3,5$ and the 4-channel reference fit. Only gate voltages corresponding to similarly low RMSE values were included in this comparison to ensure that the extracted harmonics are evaluated under equally good fits. The top panel shows the deviation for the first harmonic $v_1$, and the bottom panel for the second harmonic $v_2$. The deviation remains below $4~\%$ for $v_1$ and below $8~\%$ for $v_2$.\label{Fig4_app}}
\end{figure*}

\begin{figure*}[t!]
\centering
\includegraphics{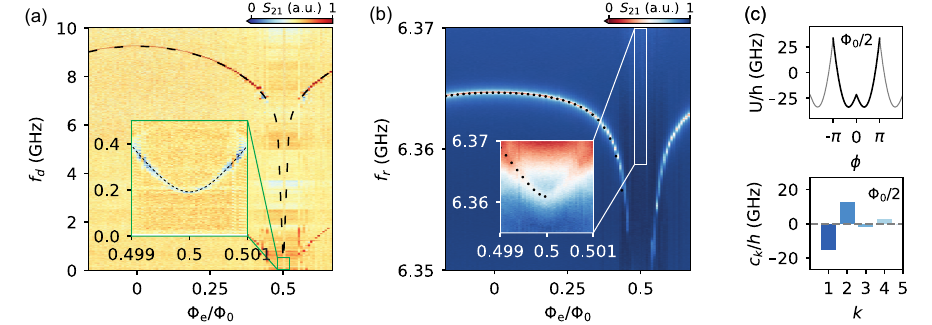}
\caption{Qubit and resonator spectroscopy at balanced first and second harmonics. (a) Qubit spectroscopy as a function of flux at $V_g=\SI{-0.7}{V}$. Dashed lines show the fit, yielding harmonics $|c_1/c_2|=1.21$. The inset shows data taken near half flux quantum. (b) Resonator spectroscopy at $V_g=\SI{-0.7}{V}$. The inset shows resonator spectroscopy taken near half flux quantum. Black dots indicate the resonator frequency simulated using parameters fitted from qubit spectroscopy.   \label{Fig2_app}}
\end{figure*}

\section{Fitting routine}\label{Sec:Fit}

To extract the physical parameters, we fix the qubit charging energy from a same-chip transmon, $E_{C_q}/h=\SI{280}{MHz}$. In fitting both the double-junction transmon and the HPQ, we assume $E_{J,1}=E_{J,2}$, as the two junctions are designed with the same nominal area. Small deviations from this symmetric case have only a weak effect on the $\phi$-dependence of the potential, as $\lambda$ remains close to unity over a wide range of junction ratios~\cite{banszerus2024voltage}.
We then select five representative datasets spanning distinct gate-voltage regimes. For each dataset, spectroscopy traces taken at multiple flux points are analyzed by identifying visible transitions (e.g., $f_{01}$, $f_{02}$, $f_{12}$, $f_{13}$, $f_{24}$) and fitting each trace with a Lorentzian to obtain transition frequencies and uncertainties. The frequency points from all five gate settings are then combined and fitted simultaneously to our Hamiltonian model. In this global fit, the three Josephson-junction–related parameters are constrained to be common across all datasets, while up to four Andreev-channel transmissions $T_i$ are allowed to vary independently with the gate voltage.

For each dataset (e.g., Fig.~\ref{Fig3_app}d), we (i) identify the transitions present, (ii) extract their frequencies via Lorentzian fits (Fig.~\ref{Fig3_app}e), and (iii) perform a least-squares fit of the collected frequency points to the model (Fig.~\ref{Fig3_app}f), with up to four Andreev channels whose transmissions $T_i$ are fit parameters.

To determine the appropriate number of conduction channels to include in the model, we compared the fit quality across models containing two to five channels. The fit quality was quantified using the root mean square error (RMSE), defined as

\begin{equation}
    \text{RMSE} = \sqrt{\frac{1}{N}\sum_{i=1}^{N} (f_i^\text{model}-f_i^\text{data})^2}
\end{equation}

where $f_i^\text{model}$ and $f_i^\text{data}$ are the best-fit and measured transition frequencies, respectively, and $N$ is the total number of fitted points. Figure~\ref{Fig4_app}(c) shows the RMSE of the fitted transition frequencies for each gate voltage. The RMSE decreases roughly an order of magnitude when increasing the number of channels from two to three, indicating that additional channels capture previously unresolved spectral features. Introducing a fourth channel further improves the fit, especially for positive gate voltages ($V_g\geq\SI{0.6}{V}$), whereas adding a fifth channel yields negligible improvement. We therefore select the least number of channels that achieve a low RMSE: a three-channel fit for datasets $V_g<\SI{0.6}{V}$ and a four-channel fit for datasets $V_g\geq\SI{0.6}{V}$.

Figure~\ref{Fig4_app}(d) compares the harmonic coefficients $v_1$ and $v_2$ obtained from the multi-channel fits ($N_\text{ch}=3,5$) to those from the four-channel reference fit. Only gate voltages yielding similarly low RMSE values were considered in the comparison, ensuring that the extracted harmonics correspond to fits of comparable quality. For gate voltages where one of the channel-number models gives a significantly larger RMSE than the others, the corresponding harmonic coefficients are not included in this comparison, since in that case the differences would reflect the poorer fit quality rather than the effect of the channel parametrization itself. The deviations remain below $4~\%$ for $v_1$ and below $8~\%$ for $v_2$. Therefore, the extracted harmonic coefficients are robust against small variations in the fitting assumptions of the number of channels.

\section{S-Sm-S junction analysis}\label{Sec:Ti}

The harmonic content of the S–Sm–S junction, extracted from fits to the qubit spectrum, is strongly gate dependent (Fig.~\ref{Fig4_app}a). As $V_g$ increases, the sum of odd coefficients ${v}_\text{odd}=\sum {v}_{2i+1}$ rises steeply, while the even sum ${v}_\text{even}=\sum {v}_{2i}$ remains smaller and grows moderately. 

To understand how each harmonic rises, Fig.~\ref{Fig4_app}a (bottom) shows individual coefficients ${v}_1,{v}_2,{v}_3,{v}_4$ normalized to their respective value at $V_g=\SI{-7}{V}$. The higher-order terms grow faster with $V_g$, although in smaller magnitude, indicating that the potential becomes increasingly nonsinusoidal as the gate opens. This trend is expected for a short junction with increasing transmission, where multiple harmonics contribute to the Josephson energy.

Fig.~\ref{Fig4_app}(b) shows the sum of the fitted transmission channels as a function of $V_g$. Small non-monotonic features with $V_g$ reflect mesoscopic behavior of the conduction channels in the nanowire. In practice, the growth of $\sum_i T_i$ shows that gate control provides a continuous knob to increase overall channel transmission and thus amplify all harmonic content.

\section{Low frequency measurement}\label{Sec:LF}

Figure~\ref{Fig2_app}a shows qubit spectroscopy versus flux, spanning both zero and half–flux-quantum points. At $\Phi_\text{e}=\Phi_0/2$, the $f_{01}$ transition drops to $f_{01}/h=\SI{191}{MHz}$, placing the \qubitacr in the low-frequency regime. The black dashed lines indicate the fit. From this fit we reconstruct the potential (Fig.~\ref{Fig2_app}c, left) and its harmonic content (Fig.~\ref{Fig2_app}c, right) at $\Phi_\text{e}=\Phi_0/2$. At this gate voltage, the first and second harmonics are comparable, with $|{c}_1/{c}_2|=1.2$.

\begin{figure*}[t!]
\centering
\includegraphics{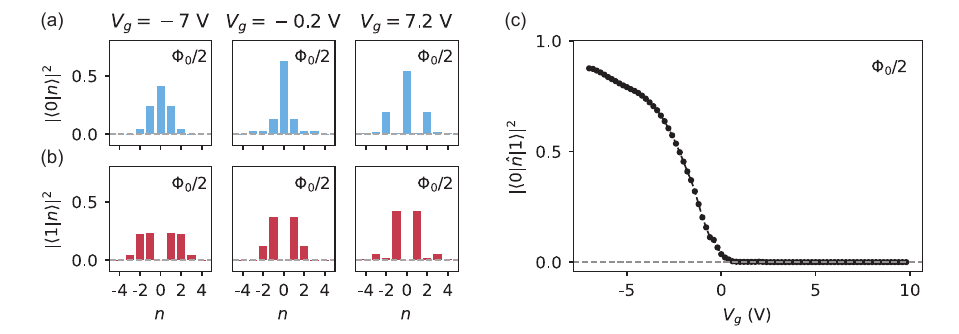}
\caption{Gate evolution of charge-parity in the HPQ wavefunctions. Charge-number wavefunction amplitudes for (a) the ground and (b) the first excited state at $\Phi_\text{e}=\Phi_0/2$. Each column corresponds to the gate voltages shown in Fig.~3 of the main text. At $V_g=\SI{-7}{V}$, the quarton-like potential produces broadly distributed charge states. At $V_g=\SI{-0.2}{V}$, (mixed-harmonic regime), $\ket{0}$ carries more even-parity weight while $\ket{1}$ carries more odd-parity weight. At $V_g=\SI{7.2}{V}$, where the $\cos(2\phi)$ term dominates, the ground state supports predominantly even charge states while the excited state supports odd charge states.\label{Fig5_app}}
\end{figure*}

\begin{figure*}[t!]
\centering
\includegraphics{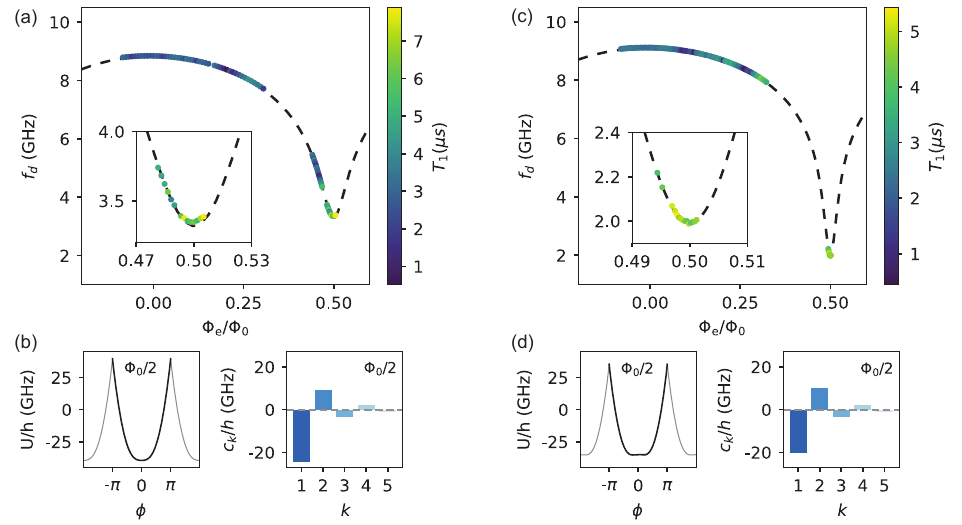}
\caption{Relaxation time versus applied flux and harmonic content at half flux quantum. (a) Markers are colored by the measured relaxation time $T_1$ at that flux point. Black dashed curves are the fitted qubit transition $f_{01}$, the inset zooms the vicinity near $\Phi_\text{e}=\Phi_0/2$. (b) Potential energy (left) and the corresponding harmonic content (right) extracted at half flux quantum. (c) Same as (a) for a second gate setting, with the dashed curves the fitted $f_{01}$ and the inset highlighting the minimum near half-flux. (d) Potential (left) and harmonic content (right) at $\Phi_0/2$ extracted from the fit in (c). \label{Fig6_app}}
\end{figure*}

\begin{figure}[t!]
\includegraphics{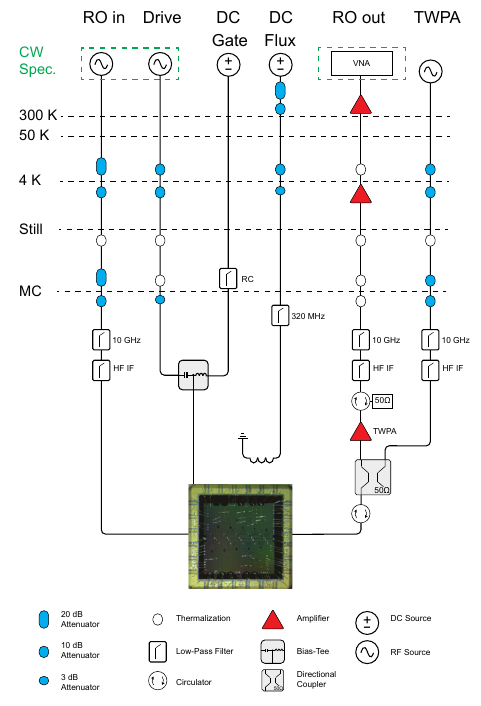}
\caption{Wiring schematic of the dilution refrigerator and device used in this study. For clarity, only one electrostatic gate-line is shown; each qubit is connected to a nominally identical line. Drive and readout lines are routed to a continuous-wave spectroscopy setup (VNA + RF source). \label{Fig1_app}}
\end{figure}

\section{Wavefunctions in charge space}\label{Sec:wvf_charge}

Fig.~\ref{Fig5_app} shows how the charge-number content of the lowest two eigenstates evolves with gate voltage at $\Phi_\mathrm{e}=\Phi_0/2$. At $V_g=-7~\mathrm{V}$ (left column), the quarton-like potential yields broadly distributed charge weights for both $\ket{0}$ and $\ket{1}$, with no strict parity preference. Moving to $V_g=-0.2~\mathrm{V}$ (middle), where the potential contains comparable even and odd harmonics, a slight parity bias emerges: the ground state moderately supports even-$n$ components, while the first excited state supports slightly more the odd-$n$ components. At $V_g=7.2~\mathrm{V}$ (right), the $\cos(2\phi)$ term dominates, and the states separate cleanly by parity—$\ket{0}$ is near pure even and $\ket{1}$ almost purely odd—with the opposite-parity components strongly suppressed.

This parity structure has direct spectroscopic consequences. Because the drive operator is proportional to $\hat n$ (odd under charge-parity), it couples states of opposite parity and suppresses matrix elements between same-parity states. The progressive even/odd segregation, therefore, explains the vanishing of transitions such as $\ket{0}\to\ket{1}$ in the high-$V_g$ regime and the reduced avoided crossings observed near half flux quantum. In short, tuning $V_g$ steers the system from mixed-parity charge superpositions to a parity-protected regime with robust selection rules.

\section{Relaxation time}\label{Sec:T1}

Fig.~\ref{Fig6_app} compares two gate settings of the same device and shows the measured relaxation time as a function of flux and frequency. In Figs.~\ref{Fig6_app}a and c, each spectroscopy point is colored by the measured $T_1$. The dashed curves are the fitted qubit transition $f_{01}$; the insets zoom into the region near $\Phi_0/2$. In both cases, $T_1$ spans $T_1\approx 1-\SI{7}{\mu s}$, with increased $T_1$ at lower frequencies, and localized dips that are consistent with parasitic two-level systems.

Figs.~\ref{Fig6_app}b and d reconstruct the potential $U(\phi)$ at $\Phi_\text{e}=\Phi_0/2$ (left) and the corresponding harmonics ${c}_k$ (right). For both gate settings, the spectrum is well described by a quarton-like potential with a dominant $\cos(\phi)$ component.

\section{Measurement setup}\label{Sec:MS}

Fig.~\ref{Fig1_app} shows the cryogenic wiring used for spectroscopy and control of the devices. From left to right it shows the readout input (RO in), qubit drive, DC gate, DC flux, readout output (RO out), and a separate TWPA pump line. Temperature stages are indicated horizontally (300 K, 50 K, 4 K, Still, and Mixing Chamber, MC). 

RF input lines (RO in and Drive) are progressively attenuated and thermalized at each stage using cryogenic attenuators (20/10/3 dB; XMA and Mini-Circuits) and thermalization points, before passing through low-pass filtering at the mixing chamber, including 10 GHz filters (Mini-Circuits) and infrared/high-frequency filters (Marki Microwave FLP series), prior to reaching the device. The DC gate and RF drive are combined through a home-built bias-tee, while the DC flux line is low-pass filtered using home-built RC filters together with Mini-Circuits VLFX-320+ filters (320 MHz cutoff) to suppress high-frequency noise and prevent qubit heating. 

The readout output chain is routed from the resonator through two cryogenic circulators/isolators at the mixing chamber (Quinstar), followed by additional filtering (Mini-Circuits and Marki Microwave) and a $\SI{50}{\Omega}$ termination. The signal is then amplified by a traveling-wave parametric amplifier (TWPA, Lincoln Laboratory), with the pump injected through a directional coupler (Krytar), and further amplified by cryogenic HEMT amplifiers at 4 K (Low Noise Factory, LNF-LNC4\_8C) and room-temperature amplifiers (Miteq) before detection at room temperature with a vector network analyzer (VNA). The TWPA pump line is independently attenuated and filtered using similar components (Mini-Circuits and Marki Microwave) before reaching the device. For clarity only one electrostatic gate line is drawn, but each qubit has a nominally identical wiring chain.

\nocite{*}

\bibliography{references}
    
\end{document}